\documentclass[useAMS]{mn2e}
\usepackage{graphicx}

\title[MONDian M/L predictions for utrafaint dSphs] {MONDian predictions for Newtonian M/L ratios for ultrafaint dSphs}

\author[R. A. M. Cort\'es, X. Hernandez] {R. A. M. Cort\'es  and X. Hernandez \\ 
Instituto de Astronom\'{\i}a, Universidad Nacional Aut\'{o}noma de M\'{e}xico,
  Apartado Postal 70--264 C.P. 04510 M\'exico D.F. M\'exico. \\
}

\date{Released 14 March 2017}

\pagerange{\pageref{firstpage}--\pageref{lastpage}} \pubyear{2015}

\begin{document}

\label{firstpage}

\maketitle

\begin{abstract}
Under Newtonian gravity total masses for dSph galaxies will scale as $M_{T} \propto R_{e} \sigma^{2}$, with $R_{e}$ the effective
radius and $\sigma$ their velocity dispersion. When both of the above quantities are
available, the resulting masses are compared to observed stellar luminosities to derive Newtonian
mass to light ratios, given a physically motivated proportionality constant in the
above expression. For local dSphs and the growing sample of ultrafaint such systems, the above results
in the largest mass to light ratios of any galactic systems known, with values in the hundreds and even thousands 
being common. The standard interpretation is for a dominant presence of an as yet undetected dark matter component.
If however, reality is closer to a MONDian theory at the extremely low accelerations relevant to such systems, $\sigma$
will scale with { stellar mass} $M_{*}^{1/4}$.
This yields an expression for the mass to light ratio which will be obtained under Newtonian assumptions of
$(M/L)_{N}=120 R_{e}(\Upsilon_{*}/L)^{1/2}$. Here we compare $(M/L)_{N}$ values from this expression to Newtonian inferences for
this ratios for the actual $(R_{e}, \sigma, L)$ observed values for a sample of recently observed ultrafaint dSphs,
obtaining good agreement. Then, for systems where no $\sigma$ values have been
reported, we give predictions for the $(M/L)_{N}$ values which under a MONDian scheme are expected once
kinematical observations become available. For the recently studied Dragonfly 44 { and Crater II systems},
reported $(M/L)_{N}$ values are also in good agreement with MONDian expectations.

\end{abstract}


\section{Introduction} \label{intro}

The standard approach towards the gravitational anomalies detected at galactic scales is to view the large measured
velocities as indirect evidence for a dominant dark matter component. Such an approach is broadly consistent with
observations and has lead to the { $\Lambda CDM$} paradigm as a unifying framework at galactic and cosmological
scales. However, over the past few years, direct detection searches and accelerator experiments have ruled out
broad classes of previously favoured dark matter candidates, driving theoretical options towards more complex
possibilities e.g. Yang et al. (2016), Szydagis et al. (2016),
CMS collaboration (2016). { The assumed driving causal entity in astrophysics and cosmology remains lacking
independent corroborating evidence.} More than the various challenges
and requirements for fine tuning of stellar feedback physics at small scales (e.g. the core/cusp problem; Moore 1994,
Flores \& Primack 1994, or the missing satellites problem; Moore et al. 1999, Boylan-Kolchin et al. 2011), it is this
persistent lack of a direct detection for the presumed dominant matter component that has increasingly led to the
exploration of a more empirical study of gravity at galactic scales.

The principal line of study in this empirical approach has been that of MOND and MONDian inspired schemes, where
observed kinematics and baryonic matter distributions are taken as the ingredients from which to infer the character
of gravity at astrophysical scales e.g. Milgrom (1984), Mendoza et al. (2011), Lelli et al. (2017). The most salient
point of these approaches is the expectation
of rotation velocities to become flat at a value consistent with the Tully-Fisher relation of $V_{rot}=(G M a_{0})^{1/4}$,
where $M$ and $a_{0}$ are the total baryonic mass of the system and the MOND acceleration scale of $1.2 \times 10^{-10}
m/s^{2}$ respectively, in the low acceleration regime where local accelerations drop below $a_{0}$. This has been
{ shown to be consistent} with rotation curves of disk galaxies of a variety of masses and galactic types, e.g.
Sanders \& McGaugh (2002), and more recently McGaugh et al. (2016) using the largest observational sample to date,
and clearly showing a one to one correlation between the distribution of baryonic matter and the resulting measured
acceleration.

In order to test for the above as a general consequence of gravity at the low acceleration scales, and not just
a peculiarity of gas dominated disks, it is important to consider also velocity dispersion supported systems in the
absence of gas. In globular clusters and dSph galaxies, the only force other than gravity is the kinematical pressure
gradients due to radial variations in stellar density and velocity dispersions. Indeed, several studies have
derived theoretical MONDian models for such pressure supported systems, and successfully compared them to
observations of Galactic globular clusters (e.g. Sanders 2012, Hernandez et al. 2017), local dSphs (e.g. Hernandez et al.
2010, Lughausen et al. 2014), ellipticals (e.g. Jimenez et al. 2013, Dabringhausen et al. 2016) and the tenuous stellar
halos surrounding external galaxies (Hernandez et al. 2013). Very recently, Durazo et al. (2017) showed that the scalings
observed in the measured projected velocity dispersion profiles of globular clusters and low rotation elliptical galaxies
from the MANGA sample also reproduce MONDian expectations.

In this context, the ultrafaint dSph galaxies become extremely relevant, as under Newtonian interpretations they are the most
'dark matter dominated' systems known (e.g. Torrealba et al. 2016b, Kim et al. 2016). In this paper we analyse a
sample of recently detected ultrafaint dSphs, to compare Newtonian mass to light ratios derived from structural and kinematical
observations, to the same quantity inferred using the MONDian scalings for low acceleration systems. The above as a critical test
of the general applicability of MONDian scalings down to the extreme low luminosity regime probed by local dSphs,
sometimes comprising as little as only a few hundred stars, e.g. Misgeld \& Hilker (2011) and references therein.

Section 2 presents the derivation of Newtonian mass to light ratios under a MONDian scenario, which are then compared
to recent observational determinations of this same quantity for a sample of recently observed ultrafaint dSphs in section 3.
In section 3 we also give predictions for the expected $(M/L)_{N}$ for a larger sample of local ultrafaint dSph systems
for which no kinematical data have yet been published. Section 4 presents our conclusions.

\section{MONDian predictions for Newtonian M/L values}

Within Newtonian gravity theory, virial equilibrium leads to a relation between the total mass of a system,
$M_{T}$, the velocity dispersion of the constituent particles, $\sigma$, and a characteristic radius, $R_{c}$, of
the form $M_{T} \propto R_{c} \sigma ^{2}$. The proportionality constant in the above expression depends on the
exact definition of $R_{c}$, the details of the radial mass profile, and on the details of the local orbital
anisotropy $\beta \equiv 1 - \sigma_t/\sigma_r$, being $\sigma_t$ the tangential and $\sigma_r$ the radial components 
of velocity dispersion, possibly radial functions, e.g. Binney \& Tremaine (1987). Indeed, it is well known that
for simple density profiles and constant anisotropy, a degeneracy appears between inferred $M_{T}$ and the assumed
value of $\beta$, for a system of known $R_{c}$ and line of sight $\sigma$. Recently however, Wolf et al. (2010)
have shown for a wide variety of Newtonian equilibrium models, that this degeneracy can be practically eliminated by
taking $R_{c}=R_{e}$, the 2D projected radius containing half the total light of a system, the effective radius, through:

\begin{equation}
M_{1/2}=4 R_{e} \frac{\sigma_{LOS}^{2}}{G}.
\end{equation}

\noindent In eq.(1) $M_{1/2}$ is half the total mass of the system and $\sigma_{LOS}$ the luminosity weighted average
value of the line of sight velocity dispersion. Assuming a constant mass to light ratio for a given system and an
observed total luminosity, $L$, the Newtonian mass to light ratio will now be given by:

\begin{equation}
(M/L)_{N} = \frac{8 R_{e} \sigma_{LOS}^{2}}{G L}.  
\end{equation}

Using eq.(1) Newtonian total dynamical masses can be accurately estimated for systems with observed ($R_{e}$ and $\sigma_{LOS}$)
values, yielding total masses with an uncertainty smaller than the errors resulting from observational confidence intervals on
$\sigma$ and $R_{e}$, this was used by Wolf et al. (2010) to infer $(M/L)_{N}$ for a range of local dSphs, and more
recently by other authors for the same purpose, e.g. Genina \& Fairbairn (2016).

If however, gravity at the low acceleration scales of local dSph galaxies behaves in a MONDian fashion,
in accordance with the Tully-Fisher relation, we can expect $\sigma_{LOS}^{2}=b(G M a_{0})^{1/2}$, with $b$
a proportionality constant of order unity, which will depend on the details of the mass profile and any
anisotropy variations, e.g. Milgrom (1984), Hernandez et al. (2013). As no models for non-spherical
equilibrium systems exist under MONDian dynamics (as local dSphs are), and no constraints are yet available
on any radial variations for $\sigma_{LOS}$ or orbital anisotropy in the local ultrafaint dSphs, where
only a generic velocity dispersion is available for some systems, in this study we take as a reasonable approximation a
constant $b=1/2$. Broadly consistent with this value of $b$ is the scaling between the observed HI velocity
and the stellar velocity dispersion of $V_{HI} = 1.33 \sigma$ found by Serra et al. (2016) for a sample of
16 rotating early type galaxies. In the absence of gas, the total { baryonic} mass of the system is given by
its stellar mass, $M_{*}$, e.g. Sanders \& McGaugh (2002). This allows to eliminate $\sigma_{LOS}$ from eq.(2) to write:

\begin{equation}
(M/L)_{N}=4 R_{e} \left( \frac{a_{0} \Upsilon_{*}}{G L} \right )^{1/2}  
\end{equation}  
 
\noindent where we have used $M_{*}=\Upsilon_{*} L$. As we are interested here in dSph galaxies, systems with no gas
and no evidence of substantial radial variations in the stellar mass to light ratio, $\Upsilon_{*}$, we shall assume
this ratio to be a constant, allowed to vary from one system to another. The above expression in astrophysical
units reads:

\begin{equation}
(M/L)_{N}=120 (R_{e}/pc) \left( \frac{\Upsilon_{*} L_{\odot}}{L} \right )^{1/2},  
\end{equation}

\noindent for $(M/L)_{N}$ in solar units. Thus, if gravity is in fact MONDian in the $a<a_{0}$ regime
valid for local ultrafaint dSphs, then we have a definitive prediction for the values for $(M/L)_{N}$
which will be inferred by studies which assume the universal validity of Newtonian gravity.

In equation (4) we already see the potential for very large $(M/L)_{N}$ values in the ultrafaint dSphs, as
this value in fact diverges as $L \rightarrow 0$. To first order, in the classical dwarfs we expect eq.(4)
to yield values in the few hundreds, for $R_{e}$ in the kpc range, $L$ of order $10^{6}$ and $\Upsilon_{*}$
of a few, as corresponds to the old stellar populations found in these systems e.g. Hernandez et al. (2000),
de Boer et al. (2014). In going to the most extreme ultrafaint dSphs, $R_{e}$ becomes of order $50 pc$, with the total
number of stars sometimes reaching only a few hundreds, again, easily yielding $(M/L)_{N}$ as inferred for these systems,
in the upper hundreds.

\section{Comparisons with observations}

\begin{figure}
\includegraphics[width=8.4cm,height=7.0cm]{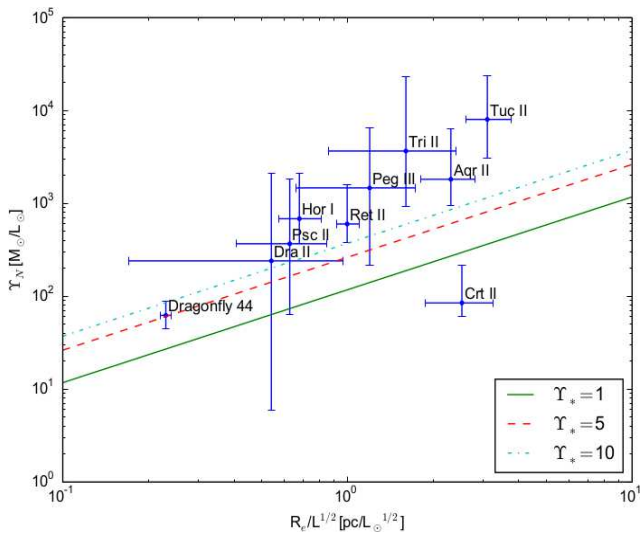}  
\caption{The figure shows inferred Newtonian dynamical mass to light ratios for a sample of recently observed
  ultrafaint dSphs, as a function of reported effective radii and luminosities, dots with error bars. The lines give
  MONDian predictions from equation (4) for the indicated range of assumed intrinsic stellar mass to light ratios.
  The recent measurements for Dragonfly 44 { and Crater II} were also included.}
\end{figure}

We begin by testing the ideas derived in the previous section by taking a sample of 8 well studied ultrafaint dSphs
for which $\sigma$, $R_{e}$ and total luminosities are available in the literature. To these 8 dSph we have added the
recently discovered Dragonfly 44 { and Crater II systems,} well measured ultra diffuse local galaxies which appear
as { two} of the most heavily dark matter dominated systems known in their luminosity ranges, van Dokkum et al. (2015),
van Dokkum et al. (2016) { and Caldwell et al. (2017)}. The observed parameters for these systems and relevant
references are summarised in table 1, { where all observed and inferred luminosities throughout our study are quoted
always in the V band.}

\begin{center}
\begin{table*}
    \begin{tabular}{| l | l | l | l | l |}
    \hline
    Galaxy  & $R_e$[pc]	& $\log$(L[L$_\odot$])	& $\sigma_\textrm{los}$[km/s]	& $\Upsilon_{\textrm{N}}$ [ $ M_{\odot}/L_{V\odot} $ ]	\\
    \hline
   	Aquarius II   & 159$\pm$24$^a$	        & 3.676$\pm$0.056$^a$	& 5.4$^{+3.4}_{-0.9}$ $^a$	& 1330$^{+3242}_{-227}$	\\
   	Crater II     & 1019$^{+250}_{-217}$$^m$	& 5.212$\pm$0.04$^m$	& 2.7$^{+1.0}_{-0.8}$$^m$	& 85$^{+133}_{-24}$\\
	Draconis II   & 19$^{+8}_{-6}$$^b$	& 3.092$\pm$0.32$^b$	& 2.9$\pm$2.1$^c$		& 501$^{+1083}_{-422}$	\\
	Dragonfly 44  & 3500$\pm$150$^d$	& 8.364$\pm$0.002$^d$	& 47$^{+8}_{-6}$$^e$		& 48$^{+21}_{-14}$		\\
	Horologium I  & 30$^{+4.4}_{-3.3}$$^f$	& 3.292$\pm$0.04$^f$	& 4.9$^{+2.8}_{-0.9}$$^g$	& 570$^{+1154}_{-112}$	\\
	Pegasus III   & 53$\pm$14$^h$		& 3.292$\pm$0.16$^h$	& 5.4$^{+3}_{-2.5}$$^h$		& 1470$^{+5660}_{-1240}$	\\
	Pisces II     & 58$\pm$7$^i$		& 3.932$\pm$0.2$^i$	& 5.4$^{+3.6}_{-2.4}$$^j$	& 370$^{+310}_{-240}$	\\
	Reticulum II  & 32$^{+1.9}_{-1.1}$$^f$	& 3.012$\pm$0.04$^f$	& 3.22$^{+1.64}_{-0.49}$$^g$     & 479$^{+904}_{-51}$		\\
	Triangulum II & 34$^{+9}_{-8}$$^b$	& 2.652$\pm$0.2$^b$	& 5.1$^{+4}_{-1.4}$$^k$		& 3600$^{+3500}_{-2100}$	\\
	Tucana II     & 165$^{+28}_{-19}$$^f$	& 3.452$\pm$0.04$^f$	& 8.6$^{+4.4}_{-2.7}$$^l$	& 1913$^{+2234}_{-950}$	\\

    \hline

    \end{tabular}
    \caption{Data for eight ultra faint dSph galaxies with available kinematical data, plus Dragonfly 44 { and Crater II}.
      $\Upsilon_N$ is calculated according
      to the method of Wolf et al. (2010). Observations from: a) Torrealba et al. (2016b), b) Laevens et al. (2015),
      c) Martin et al. (2016), d) van Dokkum et al. (2015), e) van Dokkum et al. (2016), f) Koposov et al. (2015a),
      g) Koposov et al. (2015b), h) Kim et al. (2016), i) Belokurov et al. (2010), j) Kirby et al. (2015a), k) Kirby et al.
      (2015b), l) Walker (2016) and m) Caldwell et al. (2017).}
\end{table*}
\end{center}

Total Newtonian masses for these 10 objects where then derived using eq. (1), which when divided by the observed luminosities
yields $(M/L)_{N}$. This quantity yields the y-coordinate of the points shown in figure (1), with the corresponding error bars
resulting from the error propagation of the reported observational uncertainties in $\sigma$, $R_{e}$ and $L$ for these galaxies.
{ Throughout this study we have considered the sum of reported random and systematic errors as the total error budget
on all observational quantities used, when both are available in the sources quoted.}
The x-coordinate of points in figure (1) gives the product of $R_{e} L^{-1/2}$ for each system, together with the corresponding
error bars. From eq.(4) we see that this quantity times $\Upsilon_{*}$ scales with the MONDian prediction for the Newtonian mass
to light ratios. The lines in fact give plots of eq.(4) for values of $\Upsilon_{*}=1, 5$ and $10$, bottom to top.

We see that despite the large uncertainties, the observed systems have Newtonian inferred mass to light ratios consistent with
MONDian expectations for intrinsic stellar mass to light ratios in the range $1<\Upsilon_{*}<10$. The extremely high
$(M/L)_{N}$ reported for ultrafaint dSph is hence { not in conflict with} MONDian gravity schemes in the low
acceleration regime probed by such objects. A slight trend towards requiring larger $\Upsilon_{*}$ values towards the right of
the figure is present at a very low significance level, perhaps a fortuitous occurrence due to the low number statistics and
the large error bars present, or possibly an indication of a systematic change in the $b$ parameter towards the smallest systems.

\begin{figure}
\includegraphics[width=8.4cm,height=7.0cm]{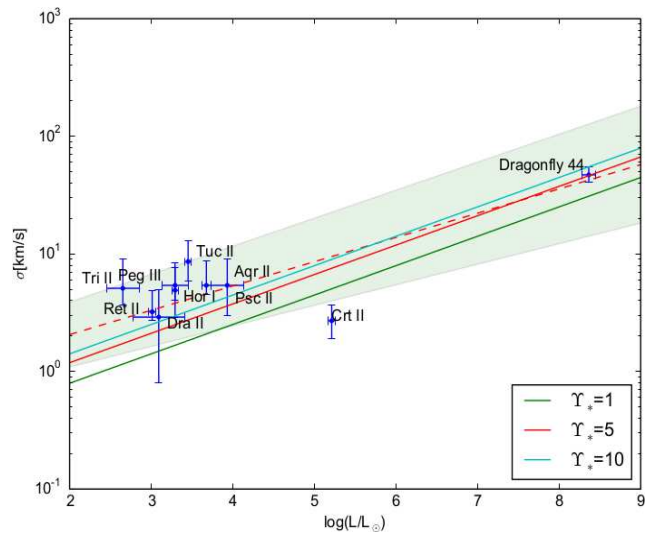}  
\caption{The dots with error bars give reported observations for the sample of ultrafaint dSphs treated, Dragonfly 44
{ and Crater II}. The dashed line gives a power law fit to the data having a slope of $0.21 \pm 0.032$, compatible with MONDian
expectations of a slope of $1/4$ { at slightly above one sigma}. The shaded region shows the $1-\sigma$ confidence
interval for this fit, including also uncertainties in the normalisation, while the solid lines give the MONDian expectations
for the shown range of intrinsic stellar mass to light ratios.}
\end{figure}

Comparing equation (1) with equation (4), it is clear that both scale linearly with $R_{e}$, which thus cancels from the test of
figure (1), which is hence really a test of the Tully-Fisher scaling of $\sigma \propto M^{1/4}$ expected under MONDian schemes
for these systems. In figure (2) we turn to a test of the above scaling by comparing two independent observed quantities
{\it a priori} uncorrelated, $\sigma$ and the total luminosity of the ultrafaint dSphs of figure 1. Notice again the
consistency of the much brighter ultradiffuse Dragonfly 44 system, the inclusion of which allows to extend the comparison
beyond the narrow range over which the ultrafaint dSphs occur, to actually allow for a meaningful power law fit to the data.
{ The same applies to slightly more than a 1 sigma level to Crater II, indeed the recent observed velocity dispersion
values for this system from Caldwell et al. (2017) are consistent with the predictions already given by McGaugh (2016)
under MOND.}

The labelled dots with error bars in figure 2 are the reported observational data for the systems in question, while the solid
lines give predictions for $\sigma$ from the $2\sigma^{2}_{LOS}=(G M a_{0})^{1/2}$ relation mentioned previously, for the observed
total luminosity measurements, assuming intrinsic stellar mass to light ratios of 1, 5 and 10, as indicated in the legend. The
dashed line gives a best fit power law to the observational data, having a slope of $0.21 \pm 0.032$, { compatible to
slightly over one sigma} with the MONDian expectation of a slope of $1/4$. The shaded region gives the $1-\sigma$ confidence
region of this fit, including uncertainties in the normalisation. We find it encouraging that the $1-\sigma$ confidence region
for the power law fit almost completely encompasses the MONDian predictions of the solid lines, for reasonable assumptions on
the stellar mass to light { ratio} of the systems treated. { That the majority of the data points appear above the dashed
line is the result of the error weighted fit applied. In fact, it is the two points outside of the ultra faint cloud which
drive the fit, which also explains the rather large confidence intervals on the resulting slope, of about 15\%. The results
shown are therefore not in conflict with a MONDian scenario, but do not in themselves force one such as a conclusion.}

The data and sources used in this first two figures are
summarised in table 1. While under standard Newtonian gravity assumptions, understanding the observed $\sigma$ values of this
systems requires the introduction of {\it ad hoc} scale dependent feedback efficiency factors, e.g. Wolf et al. (2010), under
a MONDian view point, the observed luminosities and velocity dispersions are a natural consequence of the change in gravity on
reaching the low acceleration regime.

{ Notice that the preferred M/L values we obtain are somewhat higher than typical ones
from stellar synthesis models. However, the very low metallicities and large
ages of the systems in question make the final M/L values rather uncertain.
Indeed, Caldwell et al. (2017) when reporting and modelling Crater II
consider a broad range of plausible M/L values in the V band ranging from 0.1 to 10,
as shown in the confidence band given in their figure 8. Thus, although somewhat
high, the numbers we get are within tenable values for this systems, given
uncertainties at the extreme ages and metallicities relevant.}

\begin{center}
\begin{table*}
    \begin{tabular}{| l | l | l | l | l |}
    \hline
    Galaxy	& $R_e$[pc]	& $\log$(L[L$_\odot$])	& $\sigma_\textrm{los}$[km/s]	& $\Upsilon_{\textrm{N}}$ [ $ M_{\odot}/L_{V\odot} $ ]\\
    \hline
    Columba I$^a$		& 103$\pm$25		& 3.732$\pm$0.068	& 2.83$^{+1.15}_{-0.76}$	& 284$^{+413}_{-169}$	\\
    Eridianus II$^b$	        & 277$\pm$14		& 4.772$\pm$0.12	& 5.16$^{+2.31}_{-1.50}$	& 231$^{+277}_{-121}$	\\
    Grus II$^a$			& 93$\pm$14	        & 3.492$\pm$0.088	& 2.47$^{+1.04}_{-0.69}$	& 338$^{+448}_{-188}$	\\
    Reticulum III$^a$	        & 64$\pm$24		& 3.252$\pm$0.116	& 2.15$^{+0.96}_{-0.62}$	& 307$^{+574}_{-210}$	\\
    Tucana III$^a$		& 44$\pm$6		& 2.892$\pm$0.168	& 1.75$^{+0.85}_{-0.54}$	& 319$^{+485}_{-188}$	\\
    Tucana IV$^a$		& 127$\pm$24		& 3.332$\pm$0.112	& 2.25$^{+0.99}_{-0.65}$	& 555$^{+816}_{-327}$	\\
    Tucana V$^a$		& 17$\pm$6		& 2.572$\pm$0.196	& 1.45$^{+0.74}_{-0.47}$	& 178$^{+374}_{-125}$	\\

    \hline
    \end{tabular}
    \caption{Data for seven ultra faint dSphs galaxies with no present reported kinematical data. The fourth column gives MONDian
      predicted $\sigma$ while the fifth entry shows $\Upsilon_N$ as predicted assuming the method of Wolf et al. (2010) for the MONDian
      predicted $\sigma$ values, and an $\Upsilon_*=5^{+5}_{-4}$. Observations from: a) Drlica-Wagner (2015) and b) Torrealba et al. (2016a)}

   \end{table*}
\end{center}

We end this section by giving predictions for the dynamical Newtonian mass to light ratios which can be expected from a MONDian
perspective, for a series of ultrafaint dSphs which have been identified as such (rather than as globular clusters) based
only on photometric morphological data, and for which kinematical velocity dispersion observations are not currently available.
From equation (4), we can write the expected $(M/L)_{N}$ as a function of the effective radius, the total luminosity and the
assumed intrinsic stellar mass to light ratio as:

\begin{figure}
\includegraphics[width=8.4cm,height=7.0cm]{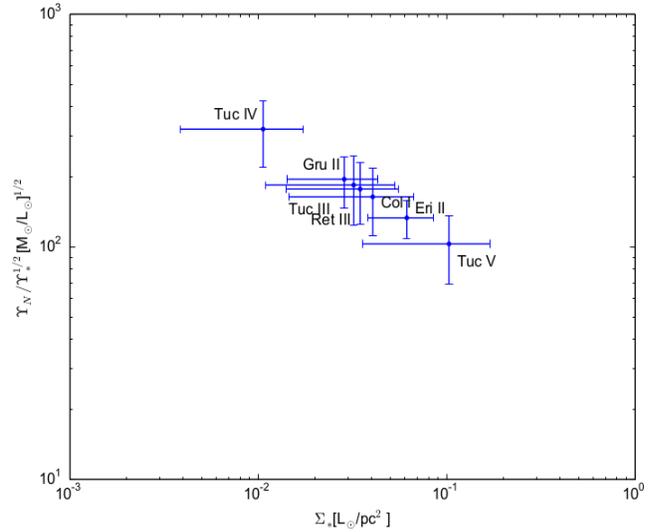}  
\caption{The figure gives MONDian predictions for Newtonian dynamical mass to light ratios for a sample of ultrafaint dSphs
currently lacking published internal dynamical measurements, but morphologically identified as galaxies, as a function of their
reported average surface brightness.}
\end{figure}

\begin{equation}
\left( \frac{M}{L} \right)_{N} = \frac{120}{(4 \pi)^{1/2}} \left(   \frac{\Upsilon_{*}}{\Sigma} \right)^{1/2} ,
\end{equation}

\noindent where we have used $\Sigma=L/(4 \pi R_{e}^2)$, the surface brightness instead of the effective radius and total luminosity.
We see at once the well known increase in the expected Newtonian mass to light ratios towards diminishing surface brightness of
MOND (e.g. Famaey \& McGaugh 2012). In figure 3 we present MONDian expectations for dynamical $(M/L)_{N}$ for a series of recently
identified ultrafaint dSphs with no currently available $\sigma$ measurements, labelled points with error bars. The horizontal
extent of the error bars is given by the reported observational uncertainties in total luminosities and effective radii, while
their vertical extent reflects a range of assumed intrinsic stellar mass to light ratios, from 1 to 10. Figure 3 hence
gives a prediction for the Newtonian dynamical mass to light ratios which we can expect from a MONDian perspective. Table 2
gives the observational data and references for the objects included in figure 3.

\section{Final remarks}

We have shown that MONDian gravity { is not inconsistent with} the observed internal dynamics for a sample of very recently
studied ultrafaint dSphs. {  We present simple first order estimates not tuned to variations which non-sphericity, departures
from isothermal conditions or the presence of orbital anisotropies would introduce. Thus,} under the MONDian perspective,
no parameters have been adjusted, as is necessary under a $\Lambda CDM$ modelling, where stellar feedback and initial baryonic
fractions have to be chosen {\it a posteriori} to explain this interesting smallest of 'galactic systems'. We give also
predictions for the expected { velocity dispersions and} $(M/L)_{N}$ values for a further series of ultrafaint systems for
which no internal dynamical observations currently exist.

\section*{acknowledgements}

We acknowledge the input of an anonymous referee as important towards reaching a more balanced and
complete final version. This work was supported in part by DGAPA-UNAM PAPIIT IN-104517 and CONACyT.

\end{document}